# Moving sand dunes

**Amelia Carolina Sparavigna**
**Dipartimento di Fisica**
**Politecnico di Torino**

In several desert areas, the slow motion of sand dunes can be a challenge for modern human activities and a threat for the survival of ancient places or archaeological sites. However, several methods exist for surveying the dune fields and estimate their migration rate. Among these methods, the use of satellite images, in particular of those freely available on the World Wide Web, is a convenient resource for the planning of future human settlements and activities.

**Introduction**
Sand dunes are hills of sand having different forms and sizes, built by the winds. The dunes can move. The first systematic study of their motion was due to Ralph Bagnold who did measurements in the Sahara Desert and experiments in wind tunnels [1]. As complex systems of particles, the dunes have attracted several studies in the past [2,3] but are also the subject of quite recent simulations [4-8]. These researches are important, because the knowledge of the dynamics of dunes can help in predicting their behaviour, in particular when the dunes can threaten some human activities.
We are often considering the fact that urban areas are cancelling the nearby nature, but there is also the reversed problem. An example is what happens at Nouakchott, the capital of Mauritania [9], because this city may be slowly covered with the sand of the moving dunes of the Sahara Desert. As the images from space are showing, the sand poses daily challenges for this city's inhabitants. The picture of Ref. [9] was coming from a Landsat satellite and shows that "hot on the city's heels are sand dunes". Another example is Chinguetti, a medieval trading centre in northern Mauritania, pole of several trans-Saharan routes [10]. Known as a holy town, rich of medieval manuscript libraries, Chinguetti is seriously threatened by the desert [11].
Besides in Africa, sand is a problem for human activities in the Middle East, China and in America [12]. According to a report prepared by the United States Geological Survey, sand dunes are becoming more mobile as the climate changes. Therefore, a monitoring of dune movement is fundamental. As discussed in [13], this monitoring can be accomplished through a number of methods, combining surface mapping with aerial and satellite imagery, GPS and LIDAR measurements. The local measurements with GPS instrumentation allow the mapping and measurements of the dune slip faces. Moreover, historical surveys of archives containing aerial and satellite imagery can be quite useful to compile a database for each dune field. Some maps can be created, such as the one proposed in the second page of Ref.13, that shows the evolution starting from 1953 of the profile of the dune field near the Grand Falls, Arizona.
In this paper, we discuss the motion of dunes. In the specific case of some barchans in Peru, we will see how they moved by comparing some maps published in the scientific literature with freely available satellites images (Google Maps). This shows that the web services are good means for monitoring the motion of dunes.

**Shape and motion of dunes**

In ancient times, the word "dune" had probably the meaning of a built-up hill [14]. These sand hills are observed having several shapes, according to the quantity of sand, the surface on which the sand moves and the blowing of the winds. Most kinds of dunes are longer on the windward side, the side where the sand is pushed up, and have a shorter side, known as the "slip face", in the lee of the wind. When the ridge is arc-shaped, this dune is a "barchan". The barchan possesses two "horns" (Fig.1 shows some barchans in Peru), the slip face inclined of approximately 35 degrees and the windward side stands at about 15 degrees [15]. We can observe isolated dunes, as those shown in Fig.1, or dune fields, where dunes coalesced. These fields that can cover regions hundreds of kilometres wide.

Using the images from satellites, we can see that the dunes in Fig.1 have their origin in a coastal area of Peru. The sand is transported inland for great distances by unidirectional winds from the oceanic shoreline, creating wide dune fields. However, besides the dunes produced by the sand that started its motion from the coastal regions, we have also huge dune fields inland in the deserts, such as in the Taklamakan Desert, or within dry lakes or dry seabeds. Let us note that sand dunes can exist in cold places too, such as in Antarctica and on Mars and Titan, the largest moon of Saturn [16].

Sand dunes move forced by wind through different mechanisms. They can move through a mechanism known as "saltation", where the particles of sand are removed from the surface and are carried by the wind, before landing back to the surface. When these particles land, they can scatter other particles and cause them to move as well. Another mechanism is present on the steep slopes of the dunes, where the sand is falling down: this is the "sand avalanche". Therefore, if we are on the sloping windward side, we can see the sand grains that jump few centimetres above the surface of the dune. At the dune's crest, the airborne sand grains fall down the steep slope as small avalanches. With strong winds, the sand particles moves in a sheet flow. This is an overland motion of the sand, having the form of a continuous layer over the soil. The mass transferred by this flow is extremely large. Then, during a strong dust storm, the dunes may move more than several meters.

As told in Ref.5, the shapes of dunes depend mainly on the amount of sand and on the yearly wind regime. The simplest type of dune is the barchan, which occurs when the wind is bowling from the same direction throughout the year. As we can see in Fig.1, barchans exist if there is not enough sand to cover the entire surface. As deduced by Bagnold, the barchans move proportionally to the wind velocity and inversely proportionally to their height. As reported in [4-8], these dunes have heights between 1.5 and 10 m, their bases are typically 40 to 150 m long and 30 to 100 m wide.

In [4-8], the researchers simulated with detailed models the creation and motion of a barchan. Let me try to show here, using a simple approach, how a crescent-shaped dune can originate. Let us assume a heap of sand with a Gaussian profile $z=H(x,y)$ in a Cartesian frame of reference, where the plane $x,y$ is the horizontal surface (the orientation is shown in Fig.2). The height of the heap is $H_o=1$ in arbitrary units. In the middle and lower part of the figure, the profile of the dune is represented by means of grey tones corresponding to $H$. The white tones correspond to $H=0$, whereas the black tones to $H=1$. The radius of the heap is $R=250$ (arbitrary units). Let us consider a wind blowing in the horizontal $x$-direction.

For each fixed value $y=Y$, the profile of the heap changes and, in particular, the dimensionless height of the corresponding profile changes. It is given by $h(Y)=\max(H(x,Y)/H_o)$. Let us subdivide the dune in several thin layers, parallel to the $x,z$ plane, having constant $y$, that is $y=Y$. Let us suppose that each layer can move with a velocity, given by the dimensionless quantity $v(Y)=1/h(Y)$, inversely proportional to the height of the layer $h(Y)$. Assuming the rate of the mass transport as a constant, we can imagine that the layer for $y=Y$ is simply shifted with this speed, without changing its profile.

We can see the evolution of the dune as a function of time in Fig.2; where the profile after a time interval defined as $n\Delta t = nR/50$ (arbitrary units), $n$ integer, is represented in grey tones. The profile of the dune is then modified in a new profile that assumes an arched shape. The "horns" have a greater speed than the central part of the dune.

This is a quite simple phenomenological approach, not considering the true mechanisms as in [4-8] But the model shows why a barchan is evolving in its horned shape. In Fig,2, in the plot for $n=0$, we have the Gaussian profile: the following evolution is shown for $n=5$ and $n=10$.

**A case in southern Peru**

Peru is rich of dune fields and wonderful barchans that can be observed from satellite. Here I show the case of the dunes in the Nazca to Tanaca area of southern Peru, based on the study reported in Ref.17 that had been done from 1959 to 1961. This quite interesting study was previously published in Spanish [18]. The researcher used air-photos of barchan dunes and plotted the rates of movement vs. dune widths. In this way, he quantified the Bagnold's work, telling that the speed of a barchan is inversely proportional to the barchan size, given by its height or width. The conclusion that the researcher is proposing in Ref.17 is that "all barchans in a given dune field, regardless of size, sweep out approximately equal areas in equal times". There is also another interesting observation of the collisions between small and large dunes. Let us suppose a larger dune in front of a small dune: the collision does not result "in destruction or absorption of the smaller dunes if the collision is a 'sideswipe'." It is observed that during the collision the dunes merge into a compound dune. After the collision, the smaller dune had passed the larger dune, retaining its approximate original size and shape.

After his studies, the researcher prepared some maps documenting the dune fields that have their origin from the sand of the Pacific coast beaches. Figure 3 shows the locations of these dune fields. We can see that the sand is blown inland by the wind having the constant direction shown by the arrows. The upper part of the figure is adapted from Fig.1 of Ref.17; in the lower part, we see tags marking the positions of some barchans. Fifty years ago, the author of Ref.17, S. Parker Gay Jr., was a geophysicist on the exploration staff of Marcona Mining. The Marcona mine is an open-pit iron mine, on the Pacific coast of Peru, 400 km southeast of Lima. In that period, a problem for this mining activity was the movement of large barchan sand dunes across the road that connected the mine with the shipping port of San Juan (see Fig.4). However, the dunes travel. Now they are far from the road, as we can see in Fig.4 on the left. On the right, we see the same image of the Google Maps, having superimposed the map of the dune set, from the Fig.2 in Ref.17. We can see the barchan dunes crossing the San Juan–Marcona highway with their positions in 1943 and 1952.

Figure 5 shows the detail of these dune set as it is today. The dunes have been scattered in this long period. Since this dune set is under constant winds and the composition of the sand is constant too, what is causing their scatter is the changing slope of the topography [17]. In fact, the dunes are moving on an irregular topography.

S. Parker Gay Jr. concluded the paper telling that the resurgent interest in aeolian geomorphology at the end of the twenty century had been very little directed to the coastal sand dunes of the Atacama Desert in Peru and Chile, "a region containing some of largest and most varied dune forms found any place in the world." I found other quantitative studies of dune fields of Peru in Ref.19. Probably other references exist.

If we observe the desert Peruvian regions using the Google Maps we see wonderful aeolian landscapes, where dunes have quite complex profiles. Besides the possibility of study and monitoring the motion of these dunes, other interesting features can be discovered. Quite recently - the paper is in press - the high-resolution Google Maps had been used to document the presence of dust devil tracks in the coastal desert of southern Peru [20]. The researcher suggests different modes of formation of dust devils related to ground surface properties. According to this reference, the location of some suitable areas on Earth, and Peru is one of these, is important for the study of dust devil tracks, promoting terrestrial analogies for Martian dust devil streaks.

**Figure Captions.**

Fig.1. When the ridge of a dune is arc-shaped the dune is a barchan. The dune has two "horns" and a slip face. In the image we see some barchans in Peru. The wind are blowing due North.

Fig.2 At the beginning, we have a heap of sand with a Gaussian profile $H(x,y)$ on the horizontal surface $x,y$. In the middle and lower part of the image, the profile of the dune has been represented by means of grey tones corresponding to $H$. The white tones correspond to $H=0$, whereas the black tones to $H=1$. The radius of the heap is $R=250$ (arbitrary units). The height is $H_o=1$ (arbitrary units). The wind is blowing in the $x$-direction. For each vertical plane with fixed $y=Y$, the profile of the heap changes. The dimensionless height of the corresponding profile is $h(Y)=\max(H(x,Y)/H_o)$. The dimensionless velocity of the layer for $y=Y$ is given by $v(Y)=1/h(Y)$, inversely proportional to the height of the profile $H(x,Y)$. We can see then the evolution of the dune as a function of time. Assuming the time step as $\Delta t = R/50$ (arbitrary units), after $n$ steps, each layer $H(x,Y)$ had moved of different distances with respect to the others. Since the speed depends on $y$, we obtain the crescent-shaped profile $h(x,y)$. The "horns" have a greater speed than the central part of the dune.

Fig.3 The upper part of the figure is adapting the Fig.1 of Ref.17 and shows the dune fields studied in that research work. In the lower part, we see tags marking the positions of some barchans, from Google and Acme Mapper.

Fig.4 - The Marcona mine is an open-pit iron mine, near the Pacific coast of Peru. A road, the black line in the lower part of the images, is connecting the mine with the shipping port of San Juan. On the left, the image from Google Maps, 2011. As reported in Ref.17, fifty years ago, a problem for the mining activity was the movement of large barchan sand dunes across the road. However, the dunes travelled. Now they are far from the road. On the right, the same image, having superimposed the dune map of Fig.2, Ref.17, where we see the dunes crossing the San Juan–Marcona highway and their positions in 1943 and 1952.

Fig.5 These are the barchans that were observed as crossing the highway fifty years ago.

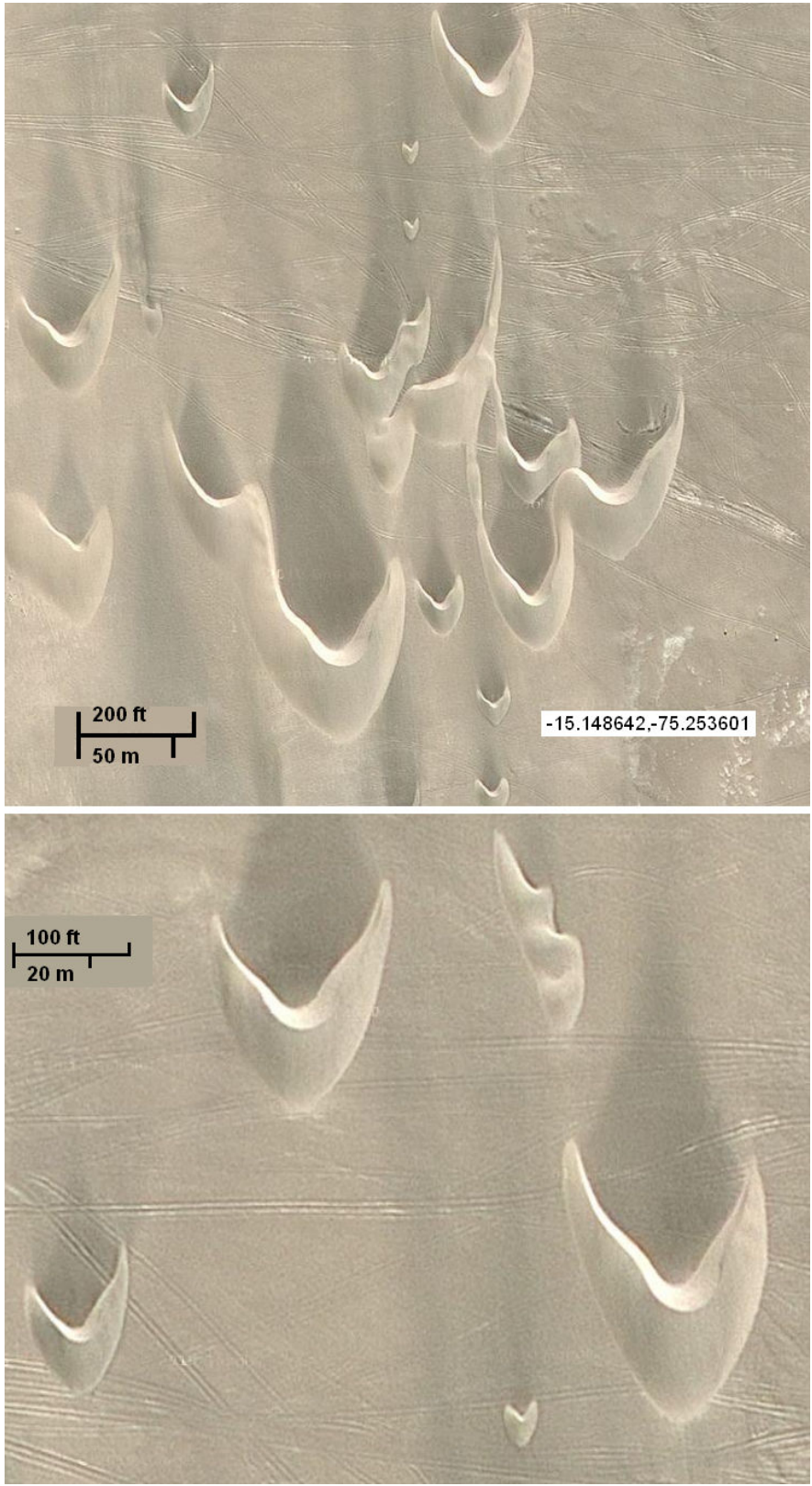

Fig.1. When the ridge of a dune is arc-shaped the dune is a barchan. The dune has two "horns" and a slip face. In the image we see some barchans in Peru. The wind are blowing due North.

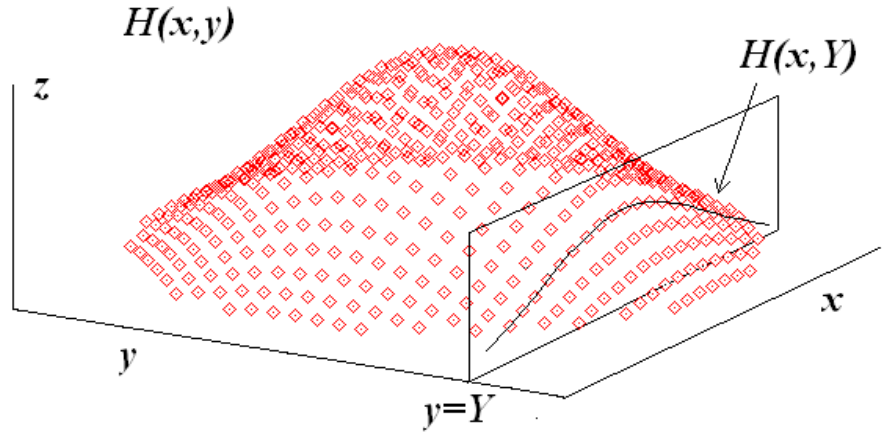

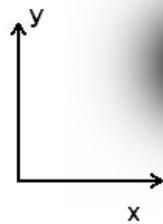

Fig.2 At the beginning, we have a heap of sand with a Gaussian profile $H(x,y)$ on the horizontal surface $x,y$. In the middle and lower part of the image, the profile of the dune has been represented by means of grey tones corresponding to $H$. The white tones correspond to $H=0$, whereas the black tones to $H=1$. The radius of the heap is $R=250$ (arbitrary units). The height is $H_o=1$ (arbitrary units). The wind is blowing in the $x$-direction. For each vertical plane with fixed $y=Y$, the profile of the heap changes. The dimensionless height of the corresponding profile is $h(Y)=\max(H(x,Y)/H_o)$. The dimensionless velocity of the layer for $y=Y$ is given by $v(Y)=1/h(Y)$, inversely proportional to the height of the profile $H(x,Y)$. We can see then the evolution of the dune as a function of time. Assuming the time step as $\Delta t = R/50$ (arbitrary units), after $n$ steps, each layer $H(x,Y)$ had moved of different distances with respect to the others. Since the speed depends on $y$, we obtain the crescent-shaped profile $h(x,y)$. The "horns" have a greater speed than the central part of the dune.

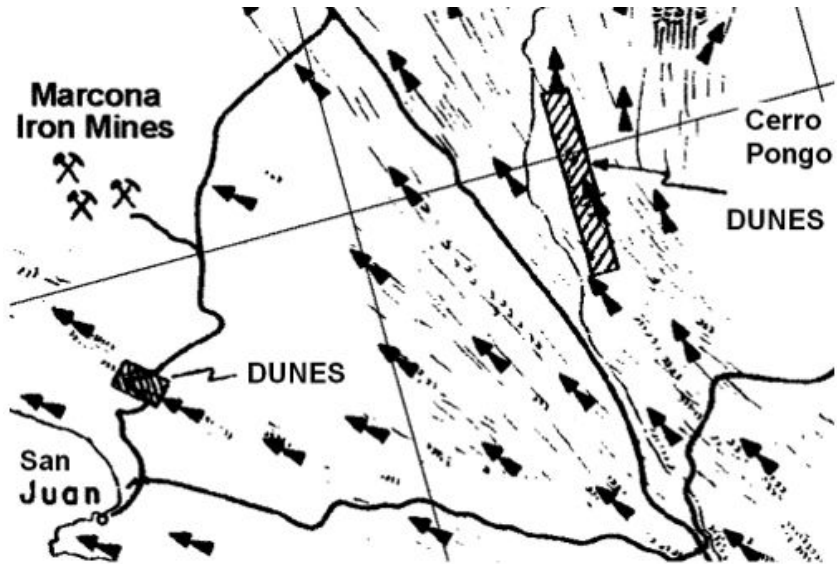

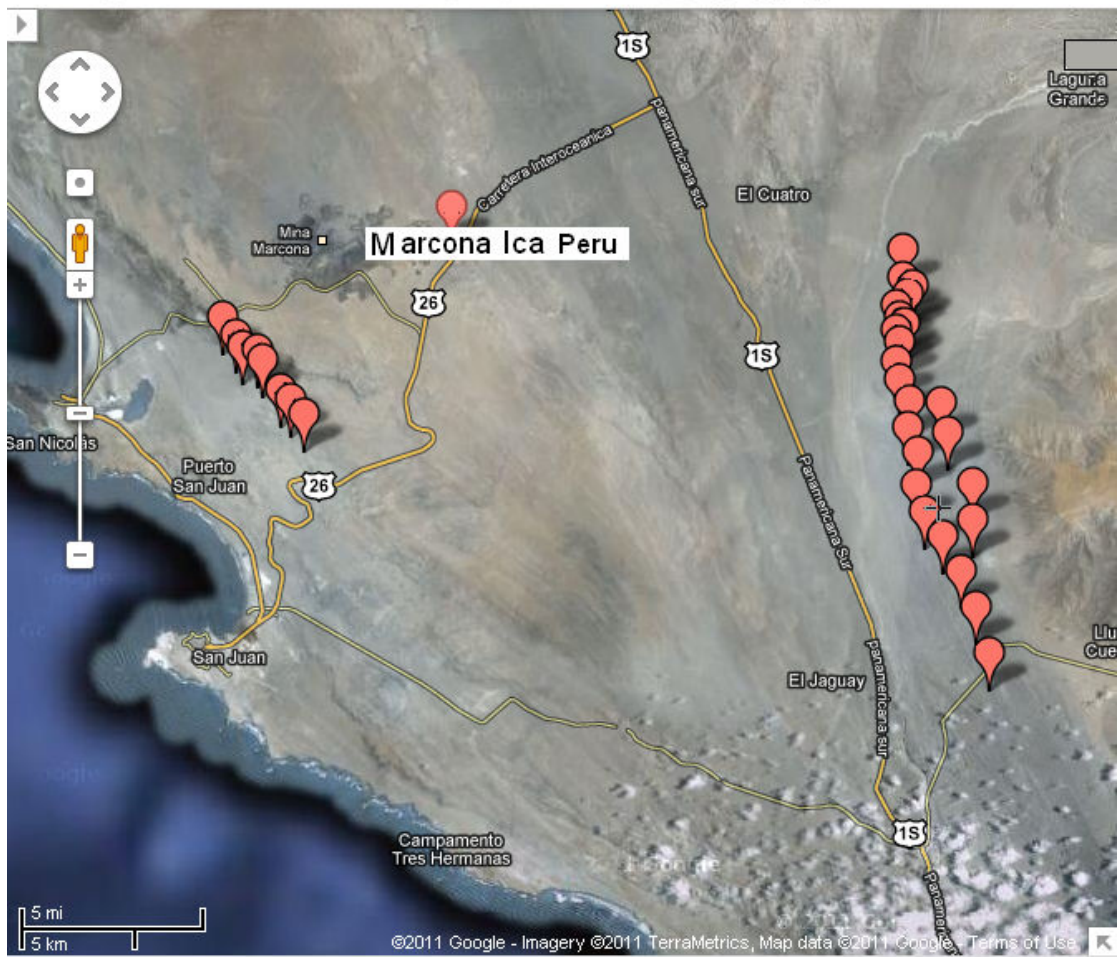

Fig.3 The upper part of the figure is adapting the Fig.1 of Ref.17 and shows the dune fields studied in that research work. In the lower part, we see tags marking the positions of some barchans, from Google and Acme Mapper.

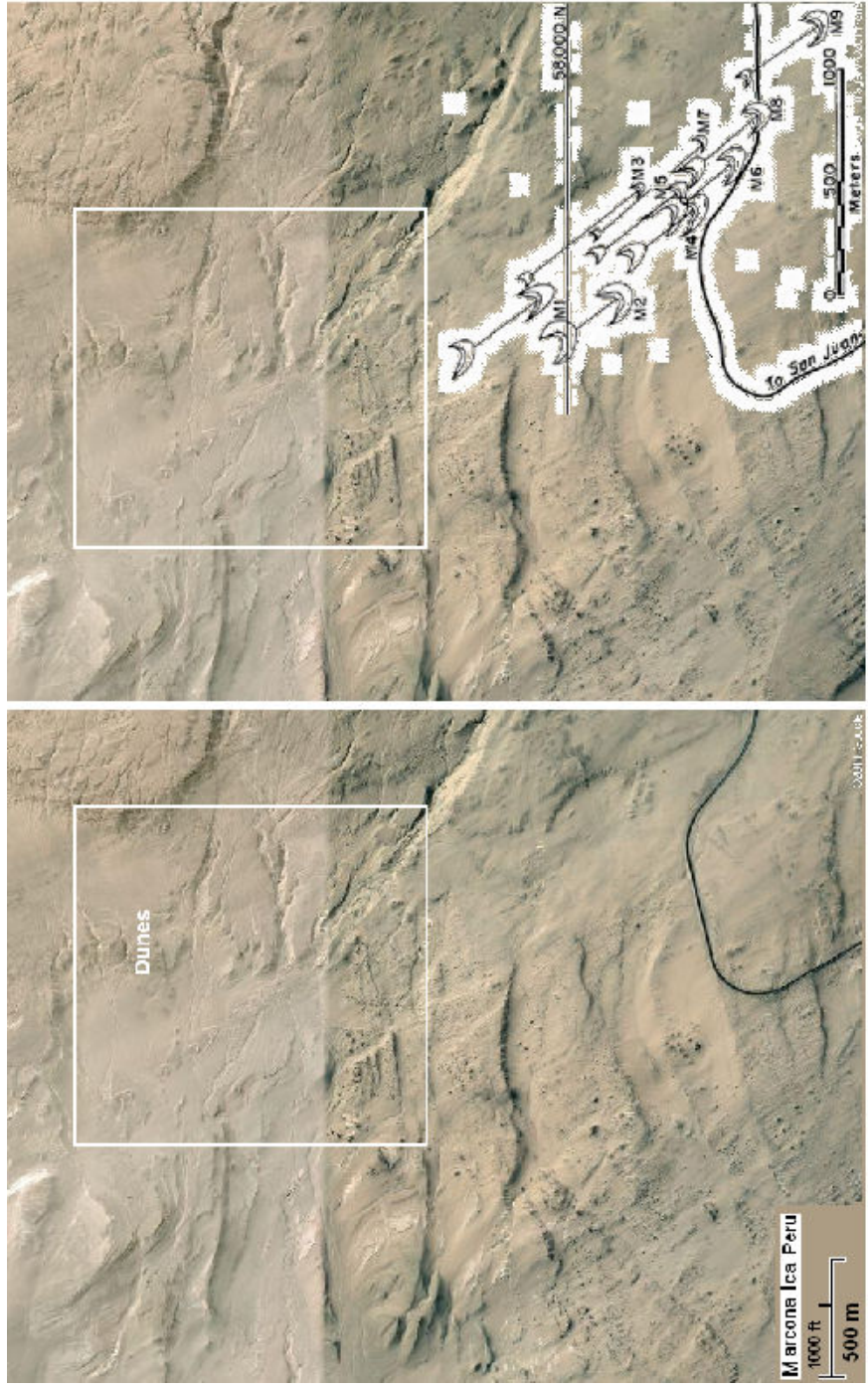

Fig.4 - The Marcona mine is an open-pit iron mine, near the Pacific coast of Peru. A road, the black line in the lower part of the images, is connecting the mine with the shipping port of San Juan. On the left, the image from Google Maps, 2011. As reported in Ref.17, fifty years ago, a problem for the mining activity was the movement of large barchan sand dunes across the road. However, the dunes travelled. Now they are far from the road. On the right, the same image, having superimposed the dune map of Fig.2, Ref.17, where we see the dunes crossing the San Juan–Marcona highway and their positions in 1943 and 1952.

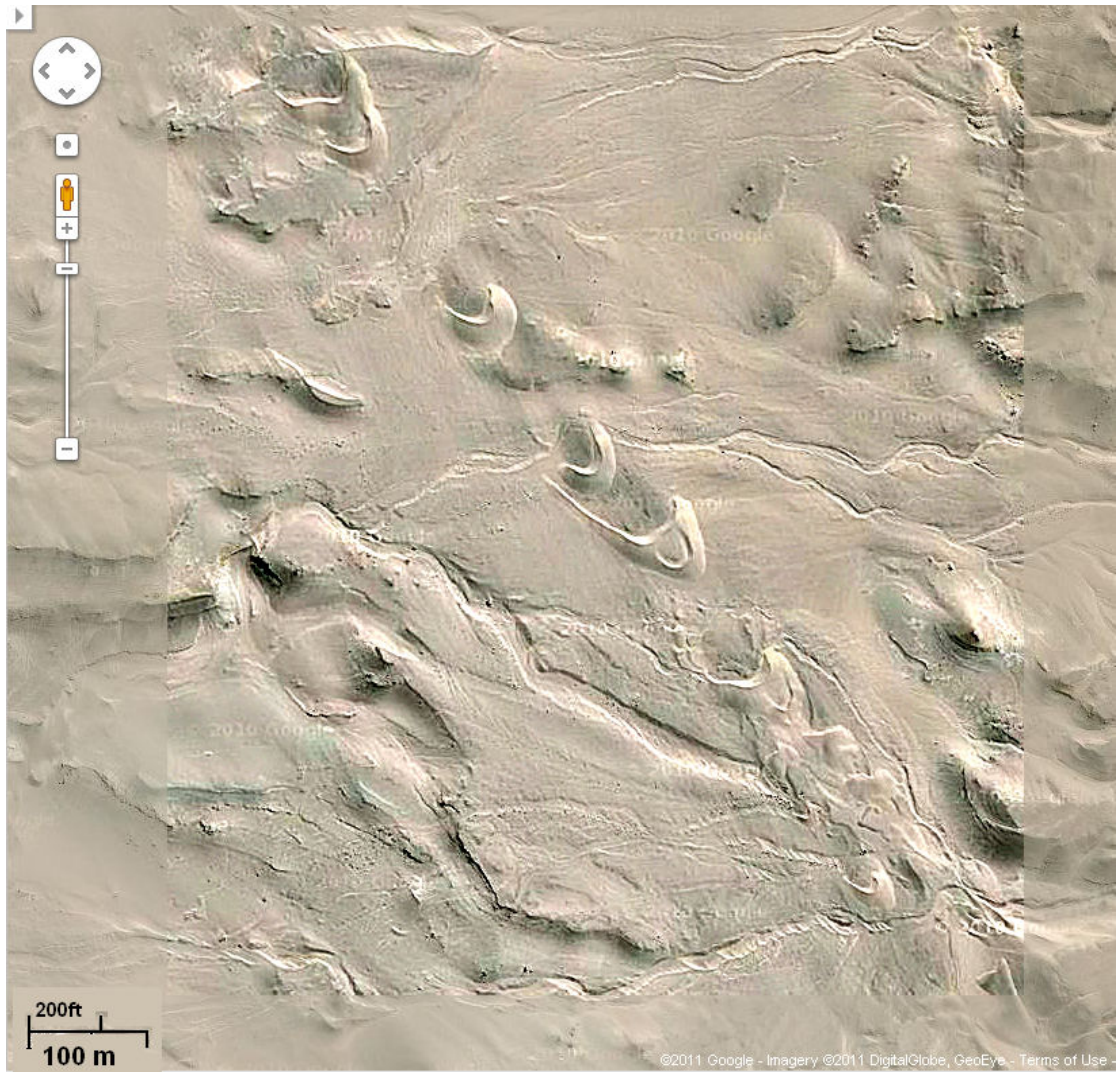

Fig.5 These are the barchans that were observed as crossing the highway fifty years ago.